\documentclass[pra,twocolumn,superscriptaddress,floatfix]{revtex4}
\usepackage{hyperref}
\usepackage{graphicx} 
\usepackage{amsmath}
\usepackage[dvips]{epsfig}
\newcommand{\beq}{\begin{equation}}
\newcommand{\eeq}{\end{equation}} 
\newcommand{\beqa}{\begin{eqnarray}}
\newcommand{\eeqa}{\end{eqnarray}}
\newcommand{\ba}{\begin{array}}
\newcommand{\ea}{\end{array}}

\usepackage{color}

\begin{document}
\title{Vortices in the supersolid phase of dipolar Bose-Einstein condensates}

\author{Francesco Ancilotto}
\affiliation{Dipartimento di Fisica e Astronomia ``Galileo Galilei''
and CNISM, Universit\`a di Padova, via Marzolo 8, 35122 Padova, Italy}
\affiliation{ CNR-IOM, via Bonomea, 265 - 34136 Trieste, Italy }

\author{Manuel Barranco}

\affiliation{Departament FQA, Facultat de F\'{\i}sica,
Universitat de Barcelona.  Av. Diagonal, 645. 08028 Barcelona, Spain}
\affiliation{Institute of Nanoscience and Nanotechnology (IN2UB),
Universitat de Barcelona, Barcelona, Spain.}

\author{Mart\'{\i} Pi}
\affiliation{Departament FQA, Facultat de F\'{\i}sica,
Universitat de Barcelona. Av. Diagonal, 645. 08028 Barcelona, Spain}
\affiliation{Institute of Nanoscience and Nanotechnology (IN2UB),
Universitat de Barcelona, Barcelona, Spain.}

\author{Luciano Reatto}
\affiliation{ Dipartimento di Fisica, Universit\`a degli Studi di Milano, 
via Celoria 16, 20133 Milano, Italy  }

\begin{abstract} 

Vortices are expected to exist in a supersolid but experimentally 
their detection can be difficult because the vortex cores 
are localized at positions where the local density is very low. 
We address here this problem by performing 
numerical simulations of a dipolar
Bose-Einstein Condensate (BEC) in a pancake confinement
at $T=0$ K
and study the effect of quantized vorticity on 
the phases that can be realized depending 
upon the ratio between dipolar and short-range interaction.
By increasing this ratio the system undergoes
a spontaneous density modulation in the form of an ordered
arrangement of multi-atom ``droplets''.
This modulated phase can be either a ``supersolid'' (SS) or a ``normal solid'' (NS). 
In the SS state
droplets are immersed in a background of low-density superfluid 
and the system has a finite global superfluid fraction
resulting in non-classical rotational inertia.
In the NS state no such superfluid background is present 
and the global superfluid fraction vanishes.
We propose here a protocol to create
vortices in modulated phases of dipolar BEC
by ``freezing'' into such phases a vortex-hosting superfluid (SF) state.
The resulting system, depending upon the interactions strengths, can be either 
a SS or a NS
To discriminate between these two possible outcome of
a ``freezing'' experiment,
we show that upon releasing of the radial harmonic confinement,
the expanding vortex-hosting SS   
shows tell-tale quantum interference effects
which display the symmetry of the vortex lattice of the originating SF,
as opposed to the behavior of the NS which
shows instead a ballistic radial expansion of the individual droplets.
Such markedly different behavior might be used to 
prove the supersolid character of rotating dipolar condensates.

\end{abstract} 
\date{\today}


\maketitle


The supersolid phase of matter has attracted large interest 
because the two symmetries that become spontaneously broken at the 
same time  seem incompatible at first sight \cite{bal10}. On one hand the 
translational symmetry is broken so that particles become localized 
with solid order. On the other hand gauge symmetry is broken and this 
leads to a condensate and to superfluid properties. Supersolidity was 
first proposed long ago for solid $^4$He \cite{and69,che70,leg70} and experiments \cite{cha13} have shown 
that solid $^4$He has indeed a number of anomalous properties but these do 
not conform to the bulk supersolid paradigm \cite{cha13,bon12}. Cold bosons turned out 
to be a more fruitful platform. 
In particular, 
a number of properties expected for a supersolid have recently been verified with 
dipolar bosons, as discussed below. In a superfluid system vortices are 
expected as excitations in addition to propagating phonon or phonon-roton ones. 
Vortices in a superfluid are quantized topological excitations directly related 
to an overall phase coherence of the system,
so the creation and detection of quantized vortices is a fundamental verification of the basic 
nature of supersolidity. 

A system made of dipolar bosons is very appealing 
in this context 
because, under suitable conditions, 
the dispersion relation of its excited states
is characterized by a roton minimum --similarly to superfluid
$^4$He-- whose  gap amplitude depends
on the relative strengths of short-range and dipolar interactions
\cite{lewe,odell}.
As the ratio between dipolar interaction and short-range
repulsion strength increases beyond a critical value, 
a spontaneous density modulation occurs
which is driven by the softening of the 
roton mode at finite momentum $k_R$,
and the resulting system shows supersolid character
\cite{wenzel,blackie_baillie3,Chomaz2018,lewe,kora,ancilotto}.
The density modulation, with wavelength $\sim 2\pi/k_R$, 
results in an ordered array of 
``droplets'', elongated in the direction of the polarization axis, 
made of many atoms each. Global phase coherence 
can be maintained between adjacent droplets
due to a low-density superfluid background that
allows tunneling of atoms from one droplet to another.
Instead, when such superfluid background is absent the
equilibrium phase consists of an ordered array of droplets
which are essentially isolated from one another (``normal solid'' phase).

Further increase in the dipolar interaction results
in the formation of 
{\it self-bound} droplets \cite{Kad16,Fer16,schmitt,santos_wachtler,blackie_baillie1,ferlaino},
with order-of-magnitude higher densities, the binding
arising from the interplay between the
two-body dipolar interactions and the effect of quantum fluctuations \cite{lima_pelster,santos_wachtler}.

Supersolid behavior occurs in the intermediate 
regime between superfluid and self-bound droplets.
We remark that the term droplet will 
be used here to indicate the individual clusters making
up the ordered structure of a modulated phase in such intermediate regime,
and will not refer to the self-bound droplet regime.

A number of theoretical studies predicted supersolid behavior in dipolar
Bose-Einstein condensates (BEC) in different geometries
\cite{bombin,cinti1,wenzel},
dipolar gases confined
in a quasi-2D pancake shaped trap \cite{blackie_baillie3,blackie_baillie1},
or in a tube \cite{ancilotto}.
The order of the superfluid-supersolid transition has been studied
in Ref. \cite{pohl}.

There is now convincing experimental evidence of supersolid (SS in the following)
behavior in dipolar gases.
Strong global phase coherence was found in
the SS realization of Ref. \cite{chomx} 
as opposed to the lack of it in the 
isolated droplet phase where no superfluid flow
is present between adjacent droplets (``normal solid'', NS in the following).
Similarly, robust phase coherence across a 
linear array of quantum droplets
was observed in Ref. \cite{bott19}.
Stable stripe modulations have been also observed
in dipolar quantum gases \cite{wenzel,tanzi}.
A partial phase coherence is suggested in Ref. \cite{tanzi},
thus indicating possible SS behavior.
The characteristic symmetry breaking of a SS phase was observed 
through the appearance of compressional oscillation 
modes in a harmonically trapped dipolar condensate \cite{tanzi_new}.
In a more recent work \cite{tanzi19} the reduction of 
the moment of inertia under slow rotation, previously predicted for
a dipolar SS \cite{stringa19}, has been measured.
The response of the dipolar SS phase has been 
studied experimentally in Ref. \cite{ferla19} where 
the out-of-equilibrium superfluid (SF in the following) flow 
across the whole system was revealed by a rapid re-establishment of 
global phase coherence after a phase-shattering excitation was applied. Instead, no such 
re-phasing was observed in the NS phase, where tunneling between adjacent droplets is suppressed.

Most of the recent evidence of supersolid 
behavior in dipolar gases is based on 
the presence of two main features of a SS system,
i.e. (i) a non-zero non-classical translational/rotational inertia \cite{sep_joss_rica}
and (ii) the appearance  of the
Nambu-Goldstone gapless mode corresponding to phase fluctuations
--besides the phonon mode associated to density fluctuations and resulting from
the translational discrete symmetry of the system.
But no evidence of another hallmark of 
superfluidity has been gathered so far, 
namely the presence of quantized vorticity \cite{stringa19}.
Experimental realization and detection of quantized vorticity would provide 
a more direct evidence of global coherence 
in the supersolid phase of a dipolar system.

In this Rapid Communication we address the properties of 
vortices in the SS phase of dipolar bosons. The 
standard way to detect vortices in a BEC is 
via visualization of the density holes in the expanding 
condensate at the positions of 
the vortex cores. This procedure is likely impossible
to apply in the SS 
state because, as shown in the following, the vortex cores are localized in 
the interstitial regions between droplets where the 
local particle density $\rho ({\bf r})$  is already very small, so that 
the additional 
depression of $\rho ({\bf r})$ at the vortex cores is almost invisible. We propose here a 
different protocol that will make possible to study 
vortices in the SS phase of dipolar BEC.
We show how a procedure based on the
``freezing'' of the vortex-hosting SF state followed by unbinding 
the system from the radial trap in the plane perpendicular 
to the polarization axis 
provides a direct access to vortical states in a 
density-modulated dipolar system: the freezing cycle allows to 
imprint a prescribed angular momentum in the modulated system
(whether it is SS or NS), while a subsequent radial expansion
allows unambiguously to discriminate if the system is in
the SS or NS state. 

We address the equilibrium structure and vortical excitations 
of a dipolar BEC confined in a pancake trap, whose short axis
is parallel to the polarization direction.

%
%
A $T=0$ dipolar BEC of atoms with
mass $m$ and magnetic moment ${\bf\mu}$  is represented by a macroscopic wave function
$\phi({\bf r})$ that obeys the extended Gross-Pitaevskii equation (eGPE) \cite{santos_wachtler}:

\begin{align}
 H\phi({\bf r})\equiv
& \left\{-\frac{\hbar^2}{2m}\nabla^2+V_t({\bf r})+g|\phi({\bf r})|^2+
\gamma(\epsilon_{dd})|\phi({\bf r})|^3+ \right.
\nonumber
\\
  & \left. \int d{\bf r'}|\phi({\bf r'})|^2V_{dd}({\bf r}-{\bf r'})
   \right\}\phi({\bf r})=\varepsilon  \phi({\bf r})
 \label{acca}
\end{align}
Here $g=4\pi\hbar^2a_s/m$, $a_s$ being the s-wave scattering length,
$V_{dd}({\bf r})=\frac{\mu_0\mu^2}{4\pi}\frac{1-3\cos^2\theta}{r^3}$
is the dipole-dipole interaction between two identical magnetic dipoles
aligned along the z axis
($\theta $ being the
angle between the vector ${\bf r}$ and the polarization direction $z$), and
$\mu_0$ is the permeability of the vacuum.
$V_t$ is the trapping harmonic potential.
The number density of the system is $\rho ({\bf r})=|\phi ({\bf r})|^2$.
The last term is the beyond-mean-field (Lee-Huang-Yang, LHY) correction \cite{lima_pelster},
where $\gamma(\epsilon_{dd})=
\frac{32}{3\sqrt{\pi}}ga^{3/2}F(\epsilon_{dd})$,
$\epsilon_{dd}=\frac{\mu_0\mu^2}{3g}$
being the ratio between the strengths of the dipole-dipole and
contact interactions, and
$F(\epsilon_{dd})=\frac{1}{2}\int_0^{\pi}d\theta
\sin\theta[1+\epsilon_{dd}(3\cos^2\theta-1)]^{5/2}$.
The chemical potential $\varepsilon $ is determined
by the normalization condition $\int |\phi ({\bf r})|^2 d{\bf r}=N$, $N$ being the total number of dipoles.

We do not find necessary to include a three-body loss term in 
our description of the dynamics because 
experimentally the
density-modulated states 
are found to be remarkably long-lived,
with a lifetime of about 150 ms \cite{chomx}, i.e. longer than the 
typical timescales investigated here.

To investigate non-zero angular momentum configurations
it is convenient to move to the fixed-cloud frame of
reference (co-rotating frame)
by imposing
a fixed value of the rotational frequency $\omega$,
i.e. we look for solutions of the equation
\begin{equation}
\left\{H \,-\omega \, \hat{L}_z \right \} \,\phi(\mathbf{r})  =  \,\varepsilon \,
\phi(\mathbf{r})
\label{eq3}
\end{equation}
where $\hat{L}_z$ is the total angular momentum operator.

We solve the above equations by propagating  
in imaginary time, if stationary states are sought, or by
propagating in real-time its time-dependent
counterpart $i\hbar \partial \phi/\partial t=H\phi$ to
simulate the dynamics of the system.
The spatial mesh spacing and time step are chosen such that 
during the time evolution excellent conservation
of the norm, total energy and angular momentum is 
guaranteed.
More details on how to solve these equations can be found in Ref. \cite{ancilotto}.

Our system is made by $N=5\times 10^5$ $^{164}$Dy 
atoms subject to harmonic trapping potential
$\omega_x=\omega_y=40\times 2\pi \,$Hz,
$\omega_z=150 \times 2\pi \,$ Hz. The shape of the
trapped dipolar gas is thus that of an oblate cloud, flattened in 
the direction  of the dipole polarization ($z$ axis).

The relative strength of the dipolar interaction 
over the short-range one is defined by the dimensionless parameter
$\epsilon _{dd}=a_{dd}/a_s$, where $a_{dd}=132$\,a$_0$ is the 
dipolar length for $^{164}$Dy atoms \cite{ferla19}.
In the following most results are for three representative values of $\epsilon _{dd}$,
i.e. $\epsilon _{dd}=1.33$, corresponding to the SF phase,
$\epsilon _{dd}=1.42$, corresponding to the SS phase,
and $\epsilon _{dd}=1.61$, corresponding to the NS phase.
The scattering lengths associated to these values are 
$a_s=99, 93$, and 82\,a$_0$, respectively.

We have preliminarily studied an extended system in the $x-y$ plane, 
i.e. we put $\omega _x=\omega _y=0$ and used periodic boundary conditions 
on a box of size $L_x=L_y= 42$ microns.  Notice that in these calculations
the confinement in the $z$ direction is still present. 
We considered different values of $\epsilon _{dd}$, and solved the Bogoliubov-de Gennes 
equations with the method described in Ref. \cite{ancilotto}.
When $\epsilon _{dd}$ is large enough the excitation spectrum 
develops a phonon-maxon-roton structure 
(see Fig. 1 in the Supplemental Material (SM)) \cite{SM}.
We find that the roton gap vanishes for $\epsilon _{dd}\sim 1.35$. 
For larger $\epsilon _{dd}$ values the homogeneous system  becomes unstable and enters 
the regime of self-modulated density, with a 
wavelength set by the wavevector at the minimum $k_R a_{ho}=1.32$, 
where $a_{ho}=\sqrt{\hbar^2/m\omega_z}=1.2112\times 10^4\,$a$_0$ is the harmonic length 
associated to the trapping potential along $z$; this 
corresponds to a wavelength $\lambda =2\pi/k_R=5.8\times 10^4$\,a$_0$.
We have also studied a vortex excitation in this extended system, with axis in 
the $z$-direction and core in the center of the box, 
i.e. we computed the lowest energy state for the wave function 
$\phi ({\bf r})=\chi ({\bf r}) exp[i\theta]$, 
where $\chi $ is a real function and $\theta $ is the angular variable in the $x-y$ plane. 
The conditions imposed on $\phi ({\bf r})$ at the boundaries 
of the calculation box are specified in Ref. \cite{Pi07}.
When $\epsilon _{dd}$ is small the vortex density profile $\chi ({\bf r})^2$ is featureless, 
vanishing quadratically at the position of the vortex core 
as in usual BEC systems. As $\epsilon _{dd}$ is increased the density profile 
develops damped oscillations near the vortex core \cite{Yi06}. 
As shown in Fig. 2 of SM \cite{SM}, the oscillations become 
more pronounced as the instability limit $\epsilon _{dd}\sim 1.35$ is approached. 
The wave vector of the oscillations is close to the roton $k_R$.
These oscillations  are similar to those predicted in superfluid $^4$He  and  
have been described as a cloud of virtual rotons \cite{reatto}. 

\begin{figure}[t]
\centerline{\includegraphics[width=1.0\linewidth,clip]{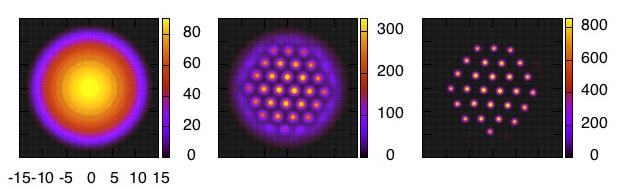}}
\caption{
Atom densities (in the plane passing through the
center of the trap) for the 
lowest energy configurations corresponding to (from left to right) 
$\epsilon _{dd}=1.33,\,1.42$,  and 1.61.
Lengths are in $\mu m$. 
The color bars show the atom densities in units of $a_{ho}^{-3}$. 
}
\label{fig1}
\end{figure}

We come now to the study of the trapped dipolar gas.  
We show in Fig. \ref{fig1} the ground-state local 
density in the  $z=0$ central plane 
for the three studied values  of $\epsilon _{dd}$
(for clarity, only the central portion of the simulation cell is shown).
For $\epsilon _{dd}=1.33$ the density is featureless and 
corresponds to that of a trapped SF cloud (left panel).
The atom density 
for $\epsilon_ {dd}=1.42$ (middle panel of Fig. \ref{fig1}) 
shows that the system spontaneously develops a
density modulation and the structure 
appears to be made of regularly arranged 
dipole clusters, elongated in the direction of the
polarization axis and immersed
in a very dilute condensate background.
The ratio $\delta $ between the density of this background 
and the maximum density of a cluster is about $\delta \sim 4.4 \times 10^{-2}$.
The spatial order is that of an approximate circular 
portion of the triangular lattice, displaying some deformation 
of the cluster lattice to minimize the energy 
in presence of the oblate trap. Besides, the density inside the 
clusters diminishes as one moves towards the border of the cloud. 
In order to avoid possible
local minima of the total energy \cite{blackie_baillie3,wenzel}
we started the imaginary-time propagation from 
the SF state shown in Fig. \ref{fig1}, changing
$\epsilon _{dd}$ to the appropriate value and adding 
some random noise (with very small amplitude) to the initial density profile.
If we start instead without this preliminary randomization,
we end up in a metastable structure similar to the one shown 
(i.e. same density modulation
wavelength) but slightly different arrangement of the droplets.
Such metastable states are also supersolid
and are likely separated by energy barriers from the 
lowest energy state shown in Fig. \ref{fig1}.


An accepted criterion which signals a supersolid behavior 
is the presence, besides of a periodic, solid-like density modulation,
of a finite non-classical rotational inertia (NCRI) \cite{leg70,sep_joss_rica}.
The latter is associated with the response of the system to a phase twist
expressed by a global superfluid fraction $f_s$ 
%
\begin{eqnarray}
f_s=1-\lim _{\omega \rightarrow 0} \frac{|\langle L_z\rangle|}{\omega I_{rb}}
\end{eqnarray}
where 
$I_{rb}=m\int \rho ({\bf r})(x^2+y^2) dx dy$ is the rigid-body
moment of inertia of the rotating cloud.
We find $f_s=1$ at $\epsilon _{dd}=1.33$ and $f_s=0.72$ at $\epsilon _{dd}=1.42$.
This verifies, respectively, the superfluid and the supersolid 
character of these two states. 
Notice that $f_s$  
should not be confused with the total superfluid fraction:
for instance, in the NS phase the droplets are individually superfluid
but do not contribute to the NCRI \cite{stringa19}.

For comparison, we show in the right panel of Fig. \ref{fig1} 
the atom density of the 
lowest energy state when $\epsilon_ {dd}=1.61$.
We find again an ordered cluster state but in this case the halo 
between clusters is practically absent ($\delta \sim 10^{-5}$).
This
state corresponds to a NS configuration, as the
calculated $f_s$ is almost zero.
Notice that the clusters appearing in the right panel of Fig. \ref{fig1} 
(which contain up to 25,000 atoms each) are not self-bound. 
When one such cluster is left isolated to evolve in time in free space (i.e. without 
any confinement) it rapidly dilutes spreading out.

It should be noticed that the theoretical 
description based on the eGPE Eq. (\ref{acca}) 
is not fully appropriate for the NS because 
by construction the same phase is assumed for all droplets 
whereas in absence of the superfluid halo the phases of 
the droplets are uncorrelated so that a more 
general many-body description is in principle needed.
Notwithstanding this, the description based on a single order parameter 
is useful because it is able to discriminate between a NS and 
a SS in terms of the character of the spectrum of 
excited states \cite{ancilotto} built upon 
the ground state and of the superfluid response that is vanishing in the NS.


\begin{figure}[t]
\centerline{\includegraphics[width=1.0\linewidth,clip]{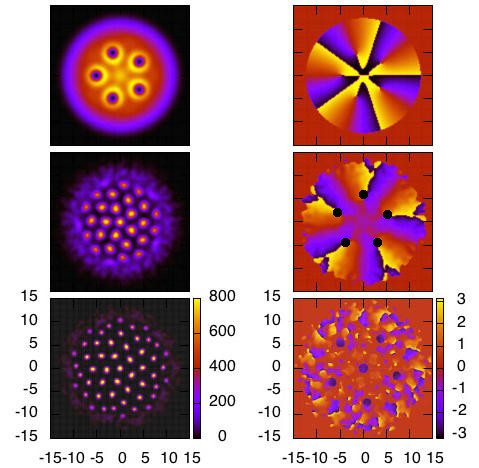}}
\caption{
Left column: atom densities of vortical states
(the quantity plotted is the integrated density $\rho_{2D}(x,y)=\int \rho (x,y,z) dz$),
with $\epsilon _{dd}=1.33,\,1.42$, and 1.61 (from top to bottom).
Lengths are in $\mu m$. The color bar shows the 
densities in units of $a_{ho}^{-2}$ and is common to all left panels.
Right column: wavefunction phase. The black dots in the middle right panel show 
the position of the five vortex cores in the SS phase.
}
\label{fig2}
\end{figure}

\begin{figure}[t]
\centerline{\includegraphics[width=1.0\linewidth,clip]{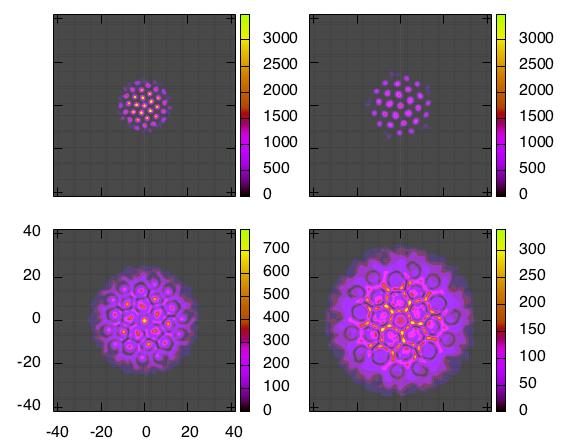}}
\caption{
Real-time expansion in the x-y plane of the SS state shown in the middle panel
of Fig. \ref{fig2}.
The four snapshots show the integrated density $\rho_{2D}$ (in units of $a_{ho}^{-2}$) 
at times $t=2.5,\,5,\,7.5,$ and 10\,ms after releasing the radial trap.
Lengths are in $\mu m$. 
}
\label{fig3}
\end{figure}

\begin{figure}[t]
\centerline{\includegraphics[width=1.0\linewidth,clip]{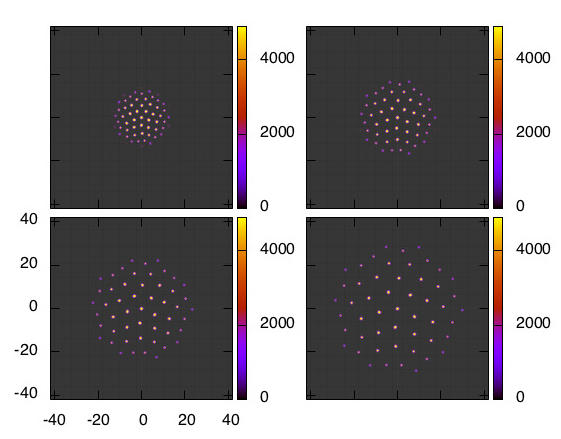}}
\caption{
Real-time expansion in the x-y plane of the NS state shown in the bottom panel
of Fig. \ref{fig2}.
The four snapshots show the integrated density $\rho_{2D}$ (in units of $a_{ho}^{-2}$)
at times $t=2.5,\,5,\,7.5,$ and 10\,ms after releasing the radial trap.
Lengths are in $\mu m$. 
}
\label{fig4}
\end{figure}

To study vortical states in the SF, SS and NS configurations described above,
we first produced a SF configurations hosting vortices
starting from the SF ground state for $\epsilon _{dd}=1.33$.
This is easily done within the co-rotating frame 
using a non-zero  $\omega$  value in Eq. (\ref{eq3}).
Since quantum mechanics
forbids rotation of oblate SF states, in order to produce vortices
the initial state has been modulated 
with a small quadrupolar distortion $exp(i \epsilon xy)$,
where $\epsilon $ is a small number;
this could be realized experimentally by using a rotating, slightly 
ellipsoidal radial trap, although other methods to produce
quantized vortices in dipolar SF have been proposed \cite{Mal11}.
Using $\omega =33$ Hz, a stationary state is reached where 
five vortices are spontaneously nucleated.
The resulting local density is shown in the  top left panel of Fig. \ref{fig2}.
One can notice the five dark spots corresponding to the vortex
cores and  the local  density oscillations outside them 
similar to those  found for a 
single vortex in the extended system. 
The phase of this state is shown in the top right panel of 
Fig. \ref{fig2}. 
The angular momentum of the rotating SF configuration
is $\langle L_z \rangle =3.33\,N \hbar$. Notice that classical
theory \cite{Hes67} predicts 
$\langle L_z\rangle/(N\hbar )=n_v (1-d^2/R^2)$
for $n_v$ linear vortices in a rotating cylinder
of radius $R$ hosting the superfluid, all with
radial distance $d$ from the axis.
Assuming the same expression, and estimating $d\sim 5.2\, \mu m$
from Fig. \ref{fig2}, one finds $R\sim 9\,\mu m$,
which is comparable with the actual size of the SF cloud shown
in the top left panel of Fig. \ref{fig2}.

We have next generated a vortical SS configuration starting
from this vortex-hosting SF state by a ``freezing'' procedure
simulated in real-time, where the scattering length
$a_s$ is linearly ramped, in 20 ms, from a value corresponding
to the SF ($\epsilon _{dd} = 1.33$) to a value $\epsilon _{dd} = 1.42$
appropriate for a SS state. We then
allow the system to further equilibrate in time for another
$80$ ms, reaching the configuration shown in the middle left
panel of Fig. \ref{fig2}. The associated phase is displayed
in the middle right panel, showing the 
presence of five singularities as expected for the presence of five vortices. 
We denote with black dots the 
positions of the singularities of the phase where the 
modulus of the wavefunction vanishes and the phase is incremented 
by $2\pi $ by turning around once.

The circulation calculated
around a closed path encircling the central portion
of the system shown in the middle panel of Fig. \ref{fig2} 
is indeed equal to 5, and 
the angular momentum is the same as in the SF state. Thus,
the original five vortices are still there, although it is impossible
to directly image the vortex cores in this SS state
because  they 
are located in the low density region between adjacent clusters in order to minimize the energy. 
An expanded view of the superfluid halo between clusters is 
shown in Fig. 3 of  SM \cite{SM}. 

The dynamics of nucleation 
of the SS phase from the SF one 
upon ramping the value of $a_s$ is shown in Ref. \cite{SM}.
The individual droplets initially nucleate
from the vortex-induced maxima of the local density,
followed by some dissociation in order to reach the correct
lattice parameter. The vortices in the initial state
seem thus to favor the nucleation of the cluster ``crystal'' and in fact 
vortices accelerate the SF-SS transition and the SS state is 
essentially reached in the ramping time of the interaction. 
We have also studied the SF-SS transition with 
the same procedure but starting from the vortex-free SF state. 
In this case we find a metastable structure very similar to the 
ground-state shown in the middle panel of Fig. \ref{fig1}.
One can also see that in the center of the trap the local order of 
the clusters is different: in the  5-vortex case 
the clusters in the center have a five-fold coordination reflecting 
the order and structure of the vortices in the SF; at variance,
in the freezing of the vortex-free SF the coordination is that of the triangular lattice,
as visible in the middle panel of Fig. \ref{fig1}.

A similar freezing cycle was performed starting
from the vortex-hosting SF state but ramping $a_s$
from the value corresponding to the SF
($\epsilon _{dd}= 1.33$) 
up to $\epsilon _{dd}= 1.61$ appropriate
for the NS state. It yielded the structure shown
in the bottom left panel of Fig. \ref{fig2}. 
The local density can be very small at positions between droplets 
and the computation of the phase is affected by large numerical errors. 
Thus in the figure the phase is put to zero 
where the local density is below a small cut-off value.



The presence of vortices in the SS state can be inferred 
from the fact that the overall cluster crystal in Fig. \ref{fig2} is 
rotating in the laboratory frame, as expected for any vortex
lattice in a superfluid system. However, also the NS is rotating
in the lab frame since it has been generated from the --angular momentum conserving-- 
freezing of a SF in a rotational state. 
Hence, this rotation is a signature of the presence
of angular momentum in the system but cannot help to distinguish a SS from a NS phase.

As a final step to distinguish between SS and NS, we perform 
a free expansion from the frozen configurations shown
in Fig. \ref{fig2} (middle and lower panels) by releasing the radial confinement (i.e. setting
$\omega _x = \omega_y = 0$). Snapshots of the ensuing real-time dynamics
are shown in Figs. \ref{fig3} and \ref{fig4}. The expanding vortex-hosting SS
state shows tell-tale quantum
interference effects where a web of density inhomogeneities develops which
reflects the 5-fold symmetry of the vortex lattice in the originating SF state, 
as opposed to the behavior of the
NS state which shows instead a ballistic radial
expansion of the individual clusters. 
Notice that during the expansion of the NS state the individual clusters
maintain their initial size without further spreading, even if they 
are not self-bound states (i.e. anyone of them, left free to evolve
in free space would expand like a gas). This effect is likely due
to a combination of the dipole-dipole repulsion between
adjacent clusters and  confinement along the perpendicular direction.
We checked that the expansion dynamics of SS and NS are robust
against small perturbations,
the expansion being almost unaffected when a small amplitude 
random noise is added to the phase of the initial state.

Remarkably, the density pattern developed during the later stages 
of the SS expansion (see last panel in Fig. 3) resembles a portion of the 
(metastable) honeycomb structure predicted for a dipolar supersolid
at larger values of the nearest-neighbor droplet distance \cite{pohl},
but with a ``defect'' represented by the 5-fold ring 
visible at the center of our structure,
which reflects the symmetry of the underlying vorticity field.
The free expansion from 
a {\it non-rotating} SS state, shown in Fig. 4 of SM \cite{SM}, also displays
similar interference effects but in this case the structure is not 
rotating, as opposed to the case of the SS obtained by freezing the vortex-hosting SF,
and the density modulations developing during the expansion 
represent a clean portion of the ``ideal''
honeycomb lattice predicted in Ref. \cite{pohl}.

Finally, one can rationalize the fact that upon expansion the droplets overlap and 
lead to interference fringes and density modulations in the SS whereas this does not happen 
in the NS as due to the presence (in the SS) or absence (in the NS) of the superfluid halo 
between droplets which acts as a glue between droplets due to the Bose statistics.
The two remarkably different behaviors uncovered by our simulations, i.e. the radial expansion 
of a rotating SS vs. the radial expansion of a rotating NS, 
allow to empirically determine the nature (SS or NS) of the 
rotating system. 

In summary, we have studied the properties of 
vortices in the SS phase of dipolar bosons
in a flattened oblate trap.
Vorticity is created in the SS by a simple 
freezing procedure starting from a vortex-hosting SF cloud
and subsequently ramping the 
scattering length characterizing the contact repulsive
interaction to 
values where one expects the spontaneous formation 
of density modulated, ordered structures.
If the freezing process leads to the SS state  
the imprinted vortices in the SF are still present in 
the SS and the cluster crystal is rotating. 
The cluster crystal is rotating also when the freezing process leads to a NS. 
To determine whether such rotating structures are a true SS or a NS lacking
global superfluid response, we simulated a free expansion 
in the $x-y$ plane:
the free expansion of the SS  generates characteristic interference patterns 
between the expanding clusters,
with symmetries that reflect the geometry of the vortex lattice
in the originating SF structure,
whereas in the case of NS 
one sees a simple ballistic expansion of the clusters driven by the
dipole-dipole repulsion between them.

\begin{acknowledgments}
One of us (F.A.) is indebted to Elena Poli for  contributing to the early stages of this work. 
This work has been supported by Grant No.  FIS2017-87801-P (AEI/FEDER, UE) (M.B., M.P.).
\end{acknowledgments}

\clearpage
\newpage

{\bf Supplemental Material}

\bigskip

\begin{figure}[!]
\centerline{\includegraphics[width=1.0\linewidth,clip]{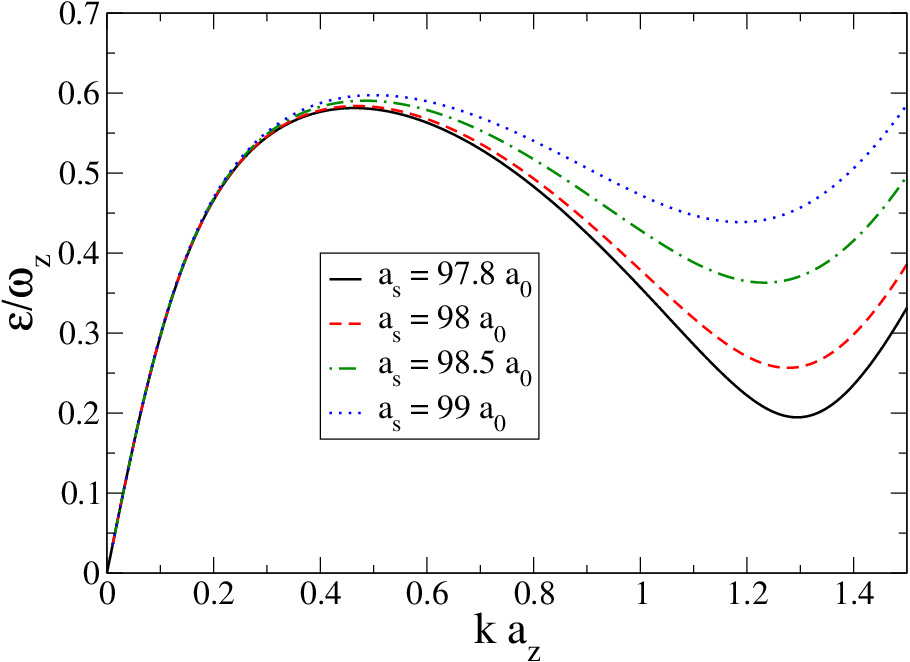}}
\caption{ Dispersion relations of the extended system for different values of $a_s$. Energies are in units of the
harmonic energy $\omega_z$ and wavevectors are in units of the inverse of the harmonic length $a_z=\sqrt{\hbar/(m\omega_z)}$.
}
\label{fig1-SM}
\end{figure}

We show in Fig. \ref{fig1-SM} the excitation spectrum of the extended system, i.e.
subject to the harmonic confinement along the z-axis but
uniform in the $x-y$ plane, for values of $a_s$ close to the 
threshold for the density instability leading to the supersolid phase.
The four curves represent the excitation spectrum for $a_s=97.8,\, 98,\, 98.5$, and 99\,a$_0$, 
which correspond to a narrow range of values of $\epsilon _{dd}$ between 
$\epsilon _{dd}= 1.35$ (corresponding to $a_s=97.8$\,a$_0$) and $\epsilon _{dd}= 1.33$ 
(corresponding to $a_s=99$\,a$_0$).
The vanishing of the roton gap occurs just below $a_s=97.8$\,a$_0$.

\begin{figure}[!]
\centerline{\includegraphics[width=1.0\linewidth,clip]{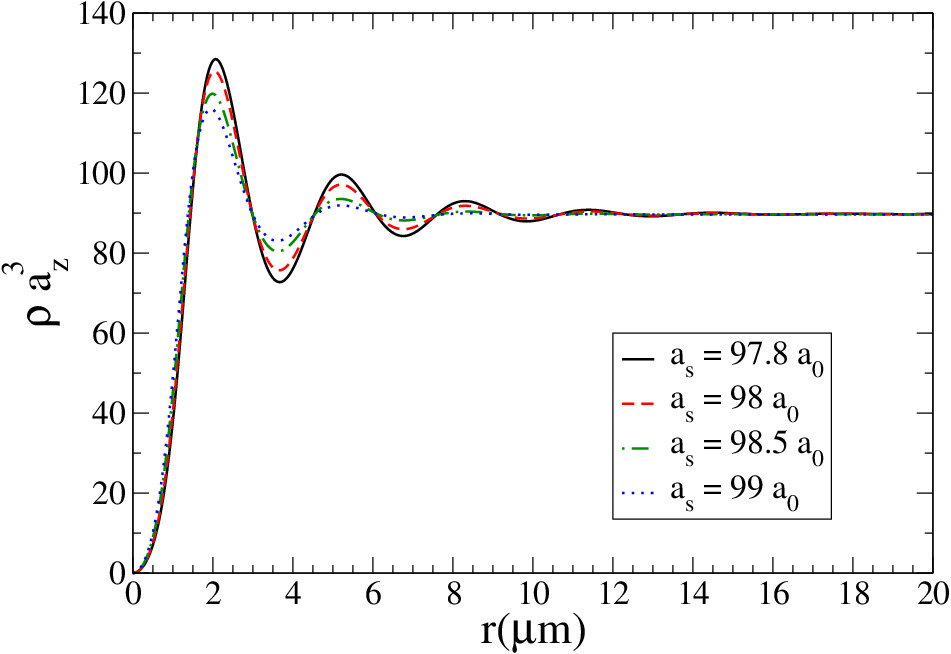}}
\caption{ Radial density profiles for a vortex in the extended system for different values of $a_s$.
}
\label{fig2-SM}
\end{figure}

We show in Fig. \ref{fig2-SM} the calculated equilibrium density profiles of singly-quantized
vortices in the extended system for the same values of $a_s$ used in Fig. \ref{fig1-SM}. 
The higher the density peaks around the vortex core, the lower the corresponding value of $a_s$.

An expanded top view of the superfuid halo between clusters in the rotating SS structure 
discussed in the main text is shown in Fig. \ref{fig3}, obtained by plotting the 
density in the $z=0$ central plane 
within a narrow range of values ($\sim 1\,\%$ of the
maximum density in the clusters). Notice the very distorted and asymmetric vortex cores,
as also found in Ref. \cite{rocc}.

\begin{figure}[!]
\centerline{\includegraphics[width=1.0\linewidth,clip]{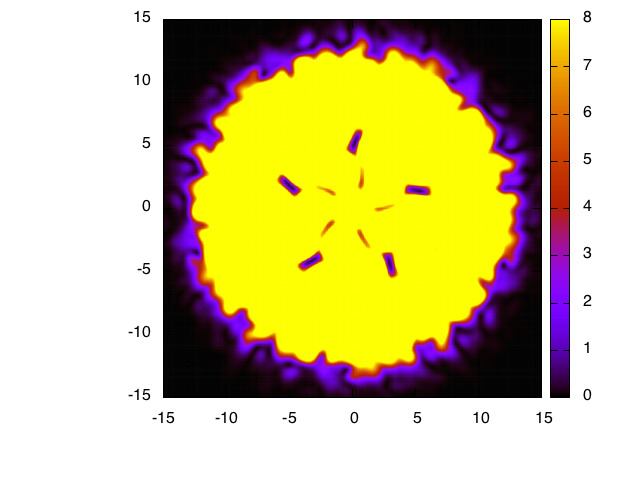}}
\caption{ Low density plot (in units of $a_{ho}^{-3}$) in the $z=0$ plane for the SS phase with vorticity.
Lengths are in $\mu m$.
}
\label{fig3}
\end{figure}

Fig. \ref{fig4} shows
the free expansion (i.e. setting
$\omega _x = \omega_y = 0$, but keeping the confinement along the
$z$-direction) of 
the SS state generated from the SF state without vortices.
The quantity plotted is the integrated density $\rho_{2D}(x,y)=\int \rho (x,y,z) dz$.
Notice the 
interference effects due to the presence of the SF background.
During the time evolution the structure appears to be non-rotating, 
as opposed to the case of the SS obtained from the SF with vortices.
During the expansion (see the bottom right panel of Fig. (\ref{fig4}))
the density shows temporarily the typical honeycomb structure which is predicted
to be a (metastable) supersolid phase of dipolar BEC 
for larger values of the ``lattice constants'' (i.e. the 
nearest-neighbor cluster distance \cite{pohl}).

\begin{figure}[!]
\centerline{\includegraphics[width=1.0\linewidth,clip]{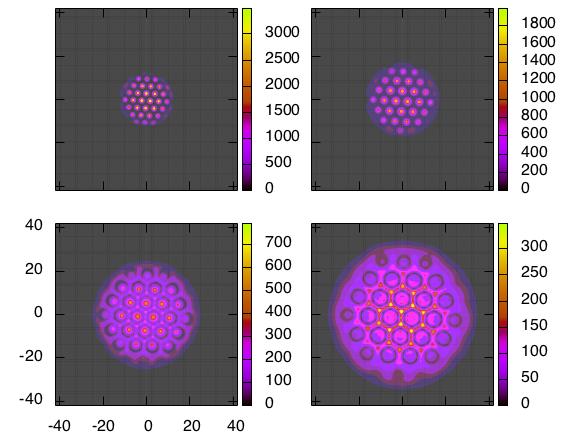}}
\caption{ Expansion of the non-rotating supersolid phase with
$\epsilon _{dd}=1.42$. 
The four snapshots show the integrated density $\rho_{2D}$ (in units of $a_{ho}^{-2}$)
at times $t=2.5,\,5,\,7.5,$ and 10\,ms after releasing the radial trap.
Lengths are in $\mu m$, and the scale is common to all panels.
}
\label{fig4}
\end{figure}

\end{document}